\documentclass[letterpaper,twocolumn,fleqn]{article} 

\usepackage{ist}

\usepackage{epsfig}
\usepackage{epstopdf}  
\usepackage{amssymb}   
\usepackage{hyperref}   
\usepackage{amsmath}    
\usepackage{mathtools}  
\usepackage{enumitem}   

\title{
\vspace{-2cm}
{\footnotesize \textnormal{\textit{Paper accepted to Media Watermarking, Security, and Forensics, IS\&T Int. Symp. on Electronic Imaging, SF, California, USA, 14-18 Feb. 2016.\\}}}\vspace{+0.5cm}\\
Deep learning is a good steganalysis tool when embedding key is reused for different images, even if there is a cover source-mismatch}

\author{Lionel PIBRE ${}^{2,3}$, J\'er\^ome PASQUET ${}^{2,3}$, Dino IENCO ${}^{2,3}$, Marc CHAUMONT ${}^{1,2,3}$\\
\small${}^1$ UNIVERSITE DE NIMES, 30 021 N\^imes Cedex 1, France\\ 
\small${}^2$ UNIVERSITE MONTPELLIER, UMR5506-LIRMM, 34 095 Montpellier Cedex 5, France \\
\small${}^3$ CNRS, UMR5506-LIRMM, 34 392 Montpellier Cedex 5, France\\
\small\texttt{\small\{lionel.pibre, jerome.pasquet, dino.ienco, marc.chaumont\}@lirmm.fr}}

\date{} 

\begin{document} 

\maketitle 

\thispagestyle{empty} 



\begin{abstract}


Since the BOSS competition, in 2010, most steganalysis approaches use a learning methodology involving two steps: feature extraction, such as the Rich Models (RM), for the image representation, and use of the Ensemble Classifier (EC) for the learning step. In 2015, Qian {\it et al.} have shown that the use of a deep learning approach that jointly learns and computes the features, was very promising for the steganalysis.

In this paper, we follow-up the study of Qian {\it et al.}, and show that in the scenario where the steganograph always uses the same embedding key for embedding with the simulator in the different images, due to intrinsic joint minimization and the preservation of spatial information, the results obtained from a Convolutional Neural Network (CNN) or a Fully Connected Neural Network (FNN), if well parameterized, surpass the conventional use of a RM with an EC.

First, numerous experiments were conducted in order to find the best "shape" of the CNN. Second, experiments were carried out in the clairvoyant scenario in order to compare the CNN and FNN to an RM with an EC. The results show more than 16\% reduction in the classification error with our CNN or FNN. Third, experiments were also performed in a cover-source mismatch setting. The results show that the CNN and FNN are naturally robust to the mismatch problem.

In Addition to the experiments, we provide discussions on the internal mechanisms of a CNN, and weave links with some previously stated ideas, in order to understand the results we obtained. We also have a discussion on the scenario "same embedding key".

\end{abstract}

\section{Introduction}
\label{sec:intro}

The state-of-the-art for steganalysis currently consists of using a two-step machine-learning methodology. 

The first step requires the extraction of features describing the image. Those features must be {\it diverse} \cite{Fridrich2011-HugoProcessDiscovery}, which means that they should capture the maximum of information modeling the image, and they also should be {\it complete} \cite{Kodovsky2008}, which means that their values should be different between a cover and a stego. The best feature set to represent an image has so far been supplied by Rich Models \cite{Fridrich2012_Rich}.

The second step consists of learning to distinguish cover model from the stego(s) model(s). Depending on the memory or computation requirements, the steganalyst can use an SVM \cite{Chih-Chung2011-SVM}, an Ensemble Classifier \cite{Kodovsky2012-EnsembleClassifiers}, or a Perceptron \cite{Lubenko_mismatch}.

When analyzing the empirical security of an embedding algorithm in a laboratory environment \cite{Ker2013_RealWorld}, we select the "clairvoyant scenario" \cite{Pevny2011-UnknownLength}, i.e. we suppose that the steganalyst knows the algorithm and the payload size that has been used by the steganograph, and has good knowledge of the cover distribution through a set of images of the same type as those used by the steganograph. 

Until 2015, in the clairvoyant scenario, the best classifier was the Ensemble Classifier \cite{Kodovsky2012-EnsembleClassifiers}. This classifier is able to treat high dimensional vectors, is easily parallelizable, and has a smaller computational complexity than SVM. Moreover, some improvements have been proposed in order to increase its efficiency, such as the use of embedding probabilities in order to better steganalyze the adaptive algorithms \cite{Denemark2014_AdaptiveSteganalysis}, tuning of false alarm probability \cite{Cogranne2014_EC_HypothesisTest}, or  treating the cover-source mismatch problem \cite{Chaumont2012-EC-FS}, where the best classifier is also the Ensemble Classifier \cite{Pasquet2014}.

Yet, in recent years, in different areas, the use of deep learning networks challenges traditional two step approaches (feature extraction, and use of a classifier) \cite{Krizhevsky2012_Convnet}. 

In the steganalysis field, Qian {\it et al.} \cite{Qian_2015_Deep} proposed, in 2015, to use deep learning to replace the traditional two step approach. In their article, Qian {\it et al.} obtained a detection percentage of only $3\%$ to $4\%$ lower\footnote{Qian {\it et al.} have probably used the "same embedding key" scenario with the use of the simulator for the embedding.} than that obtained with the Ensemble Classifier \cite{Kodovsky2012-EnsembleClassifiers}, and SRM features \cite{Fridrich2012_Rich}. The tested algorithms were HUGO \cite{Pevny2010}, WOW \cite{Holub2012_WOW}, and S-UNIWARD \cite{Holub2014}, on the BOSSbase database \cite{Bas2011-BOSS}. Those first results were encouraging since the study was only a proof of concept, and because the feature vector dimension did not compare favorably with respect to deep learning. Indeed, the dimension of the feature vector of the last convolution layer (layer 5) provides only 256 features, whereas the SRM (Spatial Rich Models) dimension of the feature vector provides 34671 \cite{Fridrich2012_Rich}.

In this article, we pursue the study of the steganalysis via deep learning in the scenario where the steganograph always uses the same embedding key for embedding in the different images and the use of the simulator for the embedding\footnote{This scenario is not a recommended scenario because this is a scenario where the steganograph weakens the security of the embedding algorithm. Note that this practical error may easily occur for example when using the C++ version of S-UNIWARD downloadable from Binghamton website.}. After many months of experiments, we obtained a reduction of more than 16\% in the classification error compared to the state of the art. We also found a network that is robust to the cover-source mismatch. 

The network we built is very different from that of Qian {\it et al.} \cite{Qian_2015_Deep}. In Section \ref{sec:cnn}, we review the major concepts of a Convolutional Neural Network. In Section \ref{sec:experiments}, we introduce the experimental settings, describe the "shape" of our best CNN, and we present steganalysis results in both scenarios: clairvoyant and  cover-source mismatch. Finally, in Section \ref{sec:discussion}, we discuss the link between the network construction and steganalysis research, and we explain our results.

\section{Convolutional Neural Network}
\label{sec:cnn}

Neural networks have been studied since the fifties. Initially, they were proposed to model the brain behavior. In computer science, especially in artificial intelligence, they have been used for 30 years for learning purposes. Until recently \cite{HintonSalakhutdinov2006b}, neural networks were considered as having a too long learning time, and as being less efficient than modern classifiers.

Recently, due to recent advances in the neural network field \cite{BengioCV13}, and to the computational power supplied by GPUs, deep learning approaches have been proposed as a natural extension of neural networks, and they are getting popular due to their high classification performance. Deep learning networks are big neural networks that can directly take data as input. In image processing, the network is directly fed with pixels. A deep learning network handles two steps at once (feature extraction and classification). Since 2006 \cite{HintonSalakhutdinov2006b}, many adjustments have been proposed to improve the robustness and reduce the computational costs.

In this paper, we recall the major concepts of a Convolutional Neural Networks (CNN), which is a deep learning network that has proved its efficiency in image classification competitions \cite{Krizhevsky2012_Convnet}, and that was used by Qian {\it et al.} for steganalysis purposes.

The learning methodology is similar to the classical one. An image database is needed, with, for each image, its label (i.e. its class). Each image is given as network input; in that case, each pixel value is taken as input of one or many neurons. The network is made of a given number of {\it layers}. A layer consists of neurons that take input values, do some computations, and then returns values that are supplied to the next layer.

\begin{equation}
F^{(0)} = \frac{1}{12}
\begin{pmatrix*}
   -1 & 2 & -2 & 2 & -1 \\
    2 & -6 & 8 & -6 & 2 \\
   -2 & 8 & -12 & 8 & -2 \\
   2 & -6 & 8 & -6 & 2 \\
   -1 & 2 & -2 & 2 & -1
\end{pmatrix*}
\label{eq:filterF0}
\end{equation}

\begin{figure*}[tb]
  \centering
  \includegraphics[width=17cm]{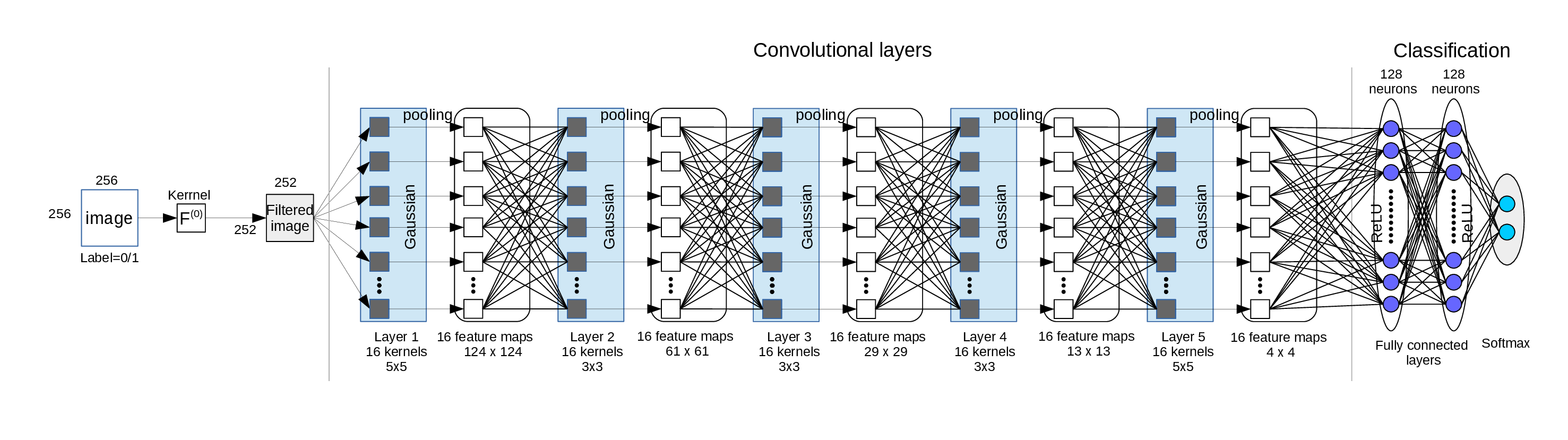}
  \caption{Qian {\it et al.} Convolutional Neural Network \cite{Qian_2015_Deep}.}
  \label{fig:qian_net}
\end{figure*}

As an illustration, Figure \ref{fig:qian_net} gives the network used by Qian {\it et al.} For this network, an image of size $256\times256$ is first filtered with a high-pass filter whose kernel is denoted $F^{(0)}$, and size is $5\times5$ (see Eq. \ref{eq:filterF0}). Note that this preliminary step is specific to the steganalysis problem. 
We observed that CNNs converge much slower without this preliminary high-pass filtering. Then, the filtered image, of size $252\times252$\footnote{The filtered image is smaller than the original image because there is no padding.}, is given to the first layer. In the Qian {\it et al.} network (see Figure \ref{fig:qian_net}), there are 5 convolution layers. 

Let us now explain more precisely what a layer is in a Convolutional Neuronal Network.

\subsection{Layer}
\label{ssec:layer}

A layer is made of neurons that take input values, do some computations, and then returns values that are supplied to the next layer. More precisely, inside a layer, computations are done in three successive steps: a {\it convolution} step (see Section \ref{ssec:convolution}), the application of an {\it activation function} (see Section \ref{ssec:activation}), and then a {\it pooling} step (see Section \ref{ssec:pooling}). Note that the outputs of a layer could be considered as a {\it set} of images. In CNN terminology, each image is named a {\it feature map}. 

In Fig. \ref{fig:qian_net}, representing the Qian {\it et al.} network, the first layer generates 16 {\it feature maps}, each of size $124\times124$. Note that this means that there are 16 filters and thus 16 convolutions which are applied to the input image of size $252\times252$. From the second to the fifth layer, there are the same three steps: convolution, activation, and pooling, but this time the convolutions are applied to all feature maps. We discuss in further detail how the convolutions and sub-sampling are fulfilled in the next subsection. 

The last convolution layer is connected to a fully connected two layer neuronal network. Then, a softmax function is connected to the outputs of the last layer in order to normalize the two outputs delivered by the network between $[0, 1]$. The softmax function gives the predicted probability of belonging to a class, while knowing the weighted outputs from the last layer. Thus, the network delivers two values as output: one giving the probability of classifying into the first class (e.g. the cover class), and the other giving the probability of classifying into the second class (e.g. the stego class). The classification decision is obtained by returning the class with the highest probability.

A Convolutional Neural Network, similar to a classical neuron network, needs a long learning time in order to tune each unknown parameter. In the case of Qian {\it et al.} network, the number of unknown parameters is close to 63 000, which indeed is not very large since convergence requires less than 2 hours with GPU programming on a Nvidia Tesla K80. Learning is achieved with the well known {\it back-propagation algorithm}. Roughly speaking, back-propagation of the error is equivalent to a gradient descent which is a well-known function optimization technique.

Network learning can thus be seen as the optimization of a function, with lots of unknown parameters, through the use of a well thought stochastic gradient descent. Due to the huge number of parameters to learn, the neural network needs a database that has a considerable number of examples in order to converge. Moreover,  the database examples must be diverse enough to obtain a good generalization of the network.

Let us now explain each step inside a layer of a Convolutional Neural Network more in detail: convolution, activation, and pooling.

\subsection{Convolution}
\label{ssec:convolution}

For a given layer and a set of {\it feature map} as input, the first processing consist of applying the convolutions. 

For the first layer, the convolution is trivial since there is only one image as input. Convolution is done between the input image and a filter. In the Qian {\it et al.} network, there are 16 filters (see Figure \ref{fig:qian_net}). Each filter leads to a filtered image. Then the second (activation function; see Section \ref{ssec:activation}), and third (pooling; see Section \ref{ssec:pooling}) steps are applied, leading to a new image named a {\it feature map}. In the Qian {\it et al.} network, there are 16 {\it feature maps} at the output of the first layer.

Formally, let $I^{(0)}$ denote the image given to the CNN (note that for the Qian {\it et al.} network, image $I^{(0)}$ is a high-pass filtered image; see Section \ref{sec:cnn} and Figure \ref{fig:qian_net}). Let $F^{(l)}_k$ denote the $k^{th}$ filter from layer $l = \{1, ..., L\}$, with $L$ beeing the number of convolutional layers, and $k\in\{1, ..., K^{(l)}\}$, with $K^{(l)}$ beeing the number of filters of the $l^{th}$ layer ($K^{(l)}$ is also the number of feature map outputs by the $l^{th}$ layer). A convolution from the first layer with the $k^{th}$ filter leads to a filtered image, denoted $\tilde{I}^{(1)}_k$, such that:

\begin{equation}
\tilde{I}^{(1)}_k = I^{(0)} \star F^{(1)}_k.
\label{eq:I1}
\end{equation}  

\begin{figure}[!htbp]
  \centering
  \includegraphics[width=8cm]{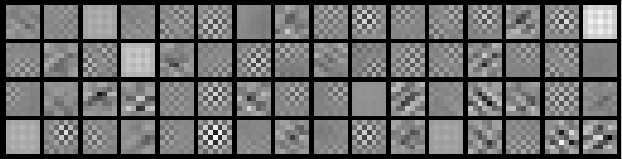}
  \caption{64 filter kernels of size $7\times7$ from the first convolutional layer obtained for our most efficient CNN.}
  \label{fig:filters_conv1}
\end{figure}

Figure \ref{fig:filters_conv1} gives an example of $64$ filter kernels obtained with the most efficient network we obtained. Note that the filter looks like oriented band-pass filters.

From the second layer to the last convolution layer, the "convolution" is less classical since there are $K^{(l-1)}$ {\it feature maps} ($K^{(l-1)}$ images) as input, denoted $I^{(l-1)}_k$ with $k=\{1, ..., K^{(l-1)}\}$. 

The "convolution" that will lead to the $k^{th}$ filtered image, $\tilde{I}^{(l)}_k$, resulting from the convolution layer numbered $l$, is in fact the sum of $K^{(l-1)}$ convolutions, such that:
\begin{equation}
\tilde{I}^{(l)}_k = \sum_{i=1}^{i=K^{(l-1)}} I^{(l-1)}_i \star F^{(l)}_{k, i},
\label{eq:I2}
\end{equation}  
with $\{F^{(l)}_{k, i}\}_{i=1}^{i=K^{(l-1)}}$ a set of $K^{(l-1)}$ filters for a given $k$ value.

\begin{figure}[!htbp]
  \centering
  \includegraphics[width=8cm]{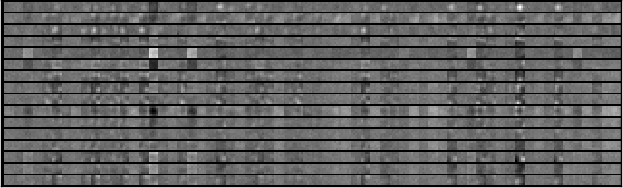}
  \caption{The $16\times64$ filter kernels of size $5\times5$ from the second convolution layer obtained for our most efficient CNN.}
  \label{fig:filters_conv2}
\end{figure}

This operation is quite unusual since each feature map is obtained by a sum of $K^{(l-1)}$  convolutions with a different filter for each convolution. 
There is no similar operation in the classical feature extraction process supplied by Rich Models \cite{Fridrich2012_Rich}, and to the best of our knowledge, in spatio-frequential decomposition. Figure \ref{fig:filters_conv2} gives an example of the 1024 filter kernels obtained by the second layer of the most efficient network we obtained.

Remember that a convolution layer is made of three steps i.e. convolution, activation, and pooling. These three consecutive steps can be summarized by looking at the link between a feature map from a layer to the previous one:
\begin{equation}
I^{(l)}_k = pool \left( f\left(b^{(l)}_k+\sum_{i=1}^{i=K^{(l-1)}} I^{(l-1)}_i \star F^{(l)}_{k,i}\right) \right),
\label{eq:layer}
\end{equation}
with $b^{(l)}_k \in \mathbb{R}$ being a scalar used for setting a bias to the convolution, $f()$ being the activation function applied pixel per pixel to the filtered image, and the pooling, $pool()$, which pools a local neighborhood (see Section \ref{ssec:pooling}).

Note that the filter kernels (also referred as weights), and the bias have to be learned and are modified during back-propagation of the error. Thus, for a layer $l\in \{1, ..., L\}$, and for a number of filters per layer, $K^{(l)}$, with a convolution kernel made of $|F^{(l)}|$ weights, there is a total number of 
 \begin{equation}
\sum_{l=1}^{l=L} K^{(l)} \times (1 + |F^{(l)}|) \nonumber
\end{equation}
unknown parameters to be tuned during back-propagation for the convolutional layer part. As an example, for the Qian {\it et al.} network, the number of parameters coming from the five convolution layers is equal to $16$ filters $\times ( 1 + 5\times 5) + 3 \times (1+3\times 3\times 16) +  (1+5\times 5\times 16)) = 13 792$ unknown parameters coming from the convolution weights and the bias (and a total of 
$63 456$ unknown parameters for the entire network if we include the fully connected layers and the softmax).

Note that the complexity computation (i.e. the number of multiplications, or the number of addition minus 1) for the convolution layers of the network is close to\footnote{The image borders are not always filtered, which actually gives a smaller number of operations.}:
\begin{equation}
\sum_{l=1}^{l=L} K^{(l)} \times |I^{(l-1)}| \times (1 + |F^{(l)}|), \nonumber
\end{equation}
with $K^{(l)}$ being the number of "convolutions" per layer, $|I^{(l-1)}|$, the size of a feature map from layer $l-1$, and $|F^{(l)}|$ the number of weights for a convolution. Note that this complexity does not take into account the {\it activation operation cost}, the {\it pooling operation cost}, and the {\it normalization operation cost}. 

Those computations are very classical and can be accelerated, for example through a fast convolution in the Fourier domain, or through the use of parallel computation. Thus, when using the network, if the size of the input image, the number of filters, the size of the filters, and the number of layers are not too big, the convolution computation cost, with the pooling and the normalization included, is more or less $O(L\times K^{(0)} \times I^{(0)} \times |F^{(2)}|)$, which is not very high. 

As an example, for the Qian {\it et al.} network, this complexity for the convolution part of the network would give more or less $5 \times 16 \times (252\times 252) \times (3 \times 3\times 16)$ additions or multiplications, which is less than 700 Mega operations. This is very small, for example, for a CPU Intel Core i7 which can compute at more than 50 GigaFLOPS, or for a GPU Nvidia Tesla K80 which can compute at more than one TeraFLOPS. Even when looking at the entire network, the computational cost is not very high. The long learning times are due to the fact that those operations have to be done on a big database, which has to be scanned many times in order that the back-propagation process does converge the network.

\subsection{Activation}
\label{ssec:activation}

Once each convolution from a convolution layer has been applied, an {\it activation} function, $f()$ (see Eq. \ref{eq:layer}), is applied to each value of the filtered image, $\tilde{I}^{(l)}_k$ (Eq. \ref{eq:I1} and Eq. \ref{eq:I2}). This function is named the activation function in reference to the notion of binary activation in the first network neuron definition. The activation function may for example be an absolute function $f(x) = |x|$, a sine function $f(x) = sinus(x)$, a Gaussian function as in the Qian {\it et al.} network $f(x) = \frac{e^{-x^2}}{\sigma^2}$, a ReLU (for Rectified Linear Units): $f(x) = max(0,x)$, etc... 

Those functions break the linearity property resulting from the linear filtering done during the convolutions. This is usually an interesting property that is exploited in the Ensemble Classifier through the majority vote \cite{Kodovsky2012-EnsembleClassifiers}, and that is also used in the Rich Models with the Min-Max features \cite{Fridrich2012_Rich}. The choice of the activation function is linked to the classification problem. For example, Qian {\it et al.} proposed to use an unusual Gaussian function. Note that the chosen function should be derivable in order to compute the back-propagation error. The derivative could be more or less computationally costly, which has an impact on the learning time. The choice of the activation function is thus often guided by this computational criteria. During the experiments, we observed that the best results were obtained with the ReLU activation function.

\subsection{Pooling}
\label{ssec:pooling}

The pooling operation consists of computing the {\it average} or {\it maximum} on a local neighborhood. In the object classification field, this operation, and especially the use of a maximum, ensures a translation invariance of the features. It was introduced in order to reduce the variance obtained during convolution. Qian {\it et al.} propose to use the average operation because the stego noise is very small. We also empirically validated this fact in our experiments. The results obtained using the average outperform the one obtained by the maximum operation.

Moreover, the pooling is coupled with a sub-sampling operation in order to reduce the size of the obtained feature map in comparison to the size of those of the previous layer. In the article of Qian {\it et al.}, there is a reduction factor of four between the feature map size of a layer and that from the previous layer. This can then be seen as a classical down-sampling with preliminary low pass filtering. This is useful to reduce the memory occupation in the GPU. Nevertheless, this step is similar to denoising, and from a signal processing point of view, it induces information loss. This pooling step does not seem interesting for a steganalysis in the scenario where the steganograph always uses the same embedding key for embedding with the simulator in the different images, and indeed we experimentally observed that suppressing the pooling operation increased the classification results by more than 8\%. In return, suppressing the pooling step gives feature maps of a constant size in all the layers, and this leads to an increase in the computational cost, and an increase in GPU memory consumption.

\section{Experiments}
\label{sec:experiments}

\subsection{The databases and learning settings used}

For the experiments, we first took the database BOSSBase v1.0 \cite{Bas2011-BOSS} consisting of 10 000 grey-level images of size $512\times512$ coming from 7 different cameras, then we split each image in four in order to obtain 40 000 images of size $256\times256$. We named this database the {\it cropped BOSSBase database}. In their article, Qian {\it et al.} also reduced the images sizes due to the GPU memory limitation. Note that they applied image resizing instead of the image cropping.

We also created a second database that we named LIRMMBase\footnote{ \includegraphics[width=1cm]{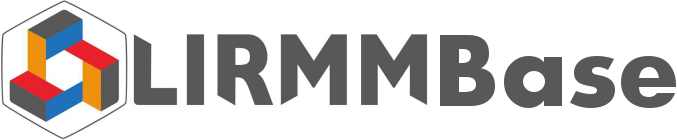} If researchers use this free-access database, they are required to cite as follows: " LIRMMBase: A database built from a mix of Columbia, Dresden, Photex, and Raise databases, and whose images do not come from the same cameras as the BOSSBase database ; For the first time used in [ref]. " L. Pibre, J. Pasquet, D. Ienco, and M. Chaumont, LIRMM Laboratory, Montpellier, France, June 2015, Website: \url{www.lirmm.fr/~chaumont/LIRMMBase.html}. [ref] Lionel Pibre, J\'er\^ome Pasquet, Dino Ienco, and Marc Chaumont, "Deep learning is a good steganalysis tool when embedding key is reused for different images, even if there is a cover source-mismatch," {\it in Proceedings of Media Watermarking, Security, and Forensics, Part of IS\&T International Symposium on Electronic Imaging, EI'2016},  San Francisco, California, USA, 14-18 Feb. 2016, 11 pages.} \includegraphics[width=1cm]{lirmmbase.png}. The LIRMMBase database consists of 1008 grey-level images, coded on 8 bits, of size $256\times256$. There are 6 cameras, none of them present in BOSSBase, and 168 images per camera. This database allows evaluation of the cover-source mismatch phenomenon. Note that readers may find different versions of the LIRMMBase on its hosting website: the color version, LIRMMBaseColor, with 15320 images, the $512\times512$ grey-level image database, LIRMMBase512x512, with 252 images, and the LIRMMBase256x256 database with 1008 grey level images. The images come from well-known databases i.e. Columbia, Dresden, Photex, and Raise databases. Note that we used the same script as that used for the BOSSBase in order to transform the RAW full resolution color images into grey-level images\footnote{The script converts the full resolution RAW color images into RGB images, then resizes the images such that the smaller side is 512 pixels long, then crops the images to $512\times512$, and finally converts them to greyscale 8-bit images; the script may be found at \url{www.lirmm.fr/~chaumont/LIRMMBase.html}.}. 

\begin{figure}[!htbp]
  \centering
  \includegraphics[width=8cm]{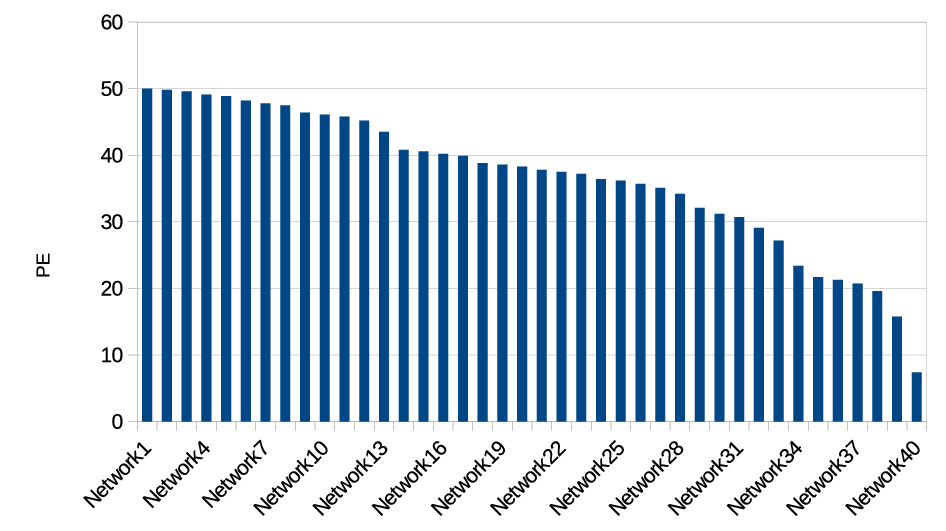}
  \caption{Probabilities of error for some of the tested CNN.}
  \label{fig:tested_CNN}
\end{figure}

In our experiments, we embeded the messages using the simulator with S-UNIWARD \cite{Holub2014} at 0.4 bits per pixel and we used the same stego key. After embedding, the database obtained from BOSSBase consisted of 80 000 images (40 000 covers and 40 000 stegos), and the database from LIRMMBase consisted of 2 016 images (1008 covers and 1008 stegos). We limited our experiments to this payload size due to the high number of computations, and experiments that we have led on the CNN. More than 40 CNN were tested with many parameter variations. Figure \ref{fig:tested_CNN} illustrates the different values of the probability of error for some of the tested CNN. 

Due to the high number of parameters, it is necessary to have a high number of iterations of the back-propagation process in order for the CNN to converge. In our experiments, the number of times the database is scanned was between 100 and 200. With a learning database consisting of 60 000 images of size $256\times256$ this leads to a learning time that is less than one day (sometimes 2 hours, sometimes more, depending on the parameters and the charge on the GPU) with the most efficient double precision GPU card on the market on June 2015, i.e. the Nvidia Tesla K80. It thus takes more than one-month of computation in order to find our "best" network. For a reproducibility of experiments, the main parameters are given: the "mini-batch" size is 128, the "moment" is 0.9, the "learning coefficient" is 0.001 for weights and 0.002 for bias, the "weight decay" is 0.004 for convolutions layers and 0.01 for the fully connected network, the "drop out" is not activated.

The paper essentially demonstrates that a CNN is more efficient than an Ensemble Classifier, or an Ensemble Classifier informed on the selection channel (adaptive steganalysis scenario) \cite{Tang2014_Adaptive, Denemark2014_AdaptiveSteganalysis} in the "same embedding key" scenario. Moreover, the paper also shows that CNN exhibits an invariance property with respect to the cover-source mismatch in this same scenario.

\begin{figure}[!htbp]
  \centering
  \includegraphics[width=8cm]{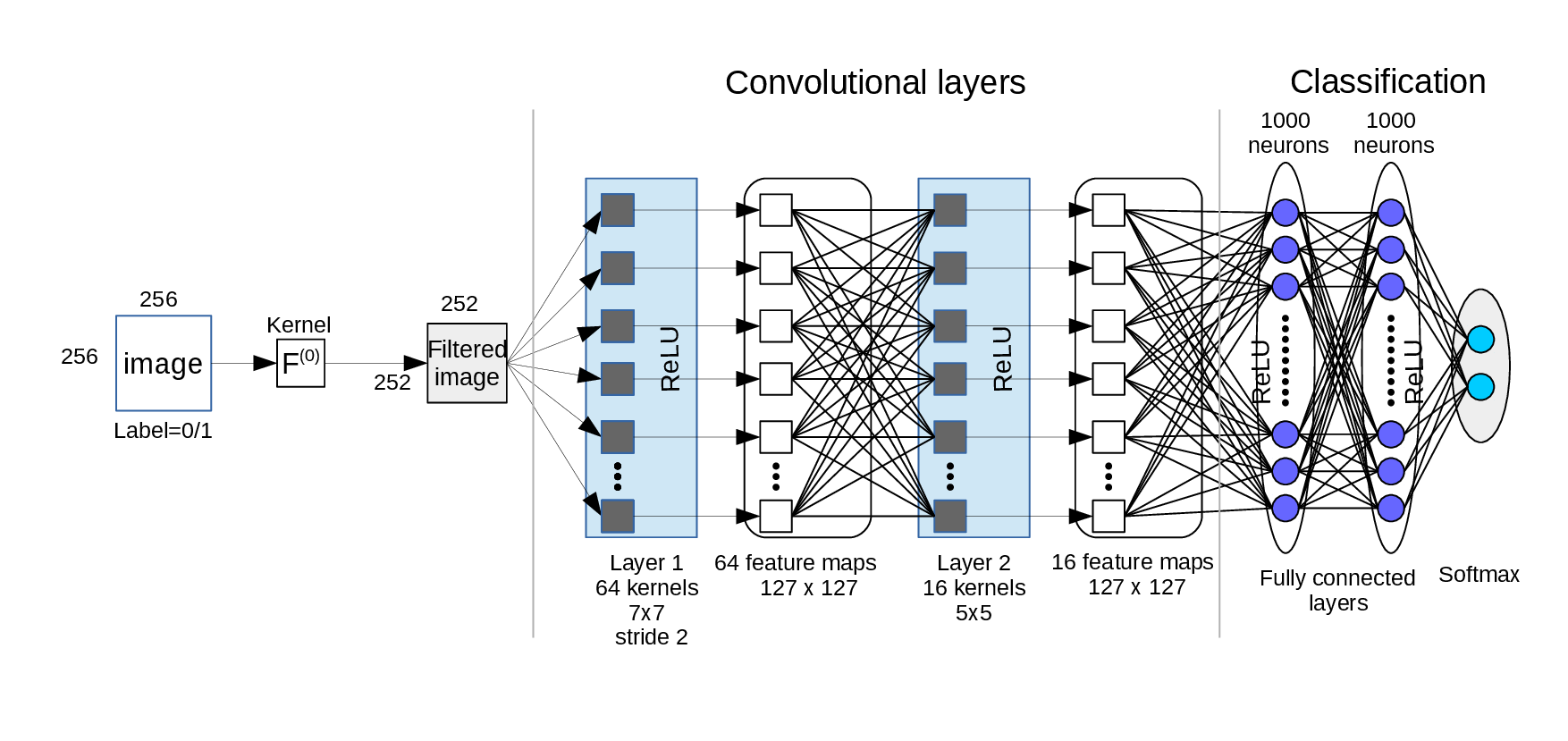}
  \caption{A Convolutional Neural Network (CNN) greater in height than in depth, and without pooling.}
  \label{fig:our_net}
\end{figure}

\subsection{The shape of our best CNN}

We tested many CNN (as shown in Figure \ref{fig:tested_CNN}), and the most efficient network we obtained consisted of only two convolutional layers, followed by a three layer fully connected network. Figure \ref{fig:our_net} illustrates this network. The input image, of size $256\times256$, is first high-pass filtered with the same filter as Qian {\it et al.} (see Eq. \ref{eq:filterF0}; and discussion in Section \ref{sec:cnn}). The size of the filtered image is thus $252\times252$. 

For the first layer, $64$ filters of size $7\times7$ are applied. Figure \ref{fig:filters_conv1} illustrates those 64 filters after a learning on 60 000 images (30000 covers and their 30000 associated stegos) from the cropped BOSSBase. As already stated, those filters seem to act as oriented band-pass filters. Note that, for memory constraints, convolution is only applied 1 pixel above 2 for lines and columns (the "stride" parameter is equal to 2) which leads to filtered images of size $127\times127$\footnote{For this first layer, the convolution with a filter size $7\times7$ on an image size of $252\times252$ virtually leads to an image size of $253\times253$, because we selected the {\it padding} option and because the {\it stride} value of 2 implies image down-sampling, which gives an image size of $127\times127$.}. Note that in comparison to the Qian {\it et al.} network, we drastically increased the number of filters in the first layer and reduced the number of CNN layers. The increase in height is one of the reasons why our network generates better results compared to the Qian {\it et al.} architecture. The increase in layer number leads to a loss of information, probably due to the negative impact of the pooling step (sub-sampling). 

After 64 convolutions, the ReLU activation function is applied  (see Section \ref{ssec:activation}); this function forces the values to be positive. Note that the use of a Gaussian activation function, as in the article of Qian {\it et al.}, does not improve the results. As already discussed, the activation breaks the linearity of the successive convolutions applied during the traverse of the convolutional layers. 

Note that we suppressed pooling because this step was counter-productive. The 64 feature maps returned by the first layer, are thus of size $127\times127$. Finally, the last process done by the layer is a normalization of the feature maps applied on each value, i.e on each position $(x,y)$ of $I^{(1)}_k$ for each $k\in\{1, ..., K^{(1)}\}$ with $K^{(1)}=64$. The normalization is done across the maps, which is useful when using unbounded activation functions such as ReLU\footnote{The normalization is done with a "local response normalization layer". {\it This enables detection of high-frequency features with a high neuron response, while damping responses that are uniformly large in a local neighborhood. This is a type of regularizer that encourages "competition" for big activities among nearby groups of neurons.} From the Cuda-convnet documentation.}:
{\small
\begin{multline}
norm(I^{(1)}_k (x,y)) = \\
  \frac{ I^{(1)}_k(x,y) } 
	{ \left(1+ \frac{\alpha}{size} \sum\limits_{k' = max(0, k - \lfloor size/2 \rfloor )}^{k' = min(K, k - \lfloor size/2 \rfloor + size )} (I^{(1)}_{k'}(x,y))^2 \right)^{\beta} }
\nonumber
\end{multline}
} with $\alpha$ and $\beta$, and $size$, the parameter set with default values\footnote{$\alpha$=0.001, $\beta$=0.75, and $size$=9.}.

The 64 feature maps of size $127\times127$ entering in the second layer are padded in order to obtain feature maps of size $131\times131$. Next, $16$ "convolutions" are applied, as explained in Section \ref{ssec:convolution}. Equation \ref{eq:I2} recalls this particular convolution step, where each sum of convolutions can be seen as the research of local signals in each of the feature maps through computation of the correlations between the researched signals and the feature maps. This step, and the normalization steps, probably explain the robustness to the cover-source mismatch that we observed during the experiments. 

Figure \ref{fig:filters_conv2} illustrates those $16\times64$ filters after a learning on 60 000 images (30000 covers and their 30000 associated stegos) from the cropped BOSSBase. ReLU activation is employed after applying the convolutions with those filters, and then the normalization is performed again. This leads to 16 feature maps of size $127\times127$ each. Note that concatenating all feature maps values leads to a feature vector consisting of 258 064 features,
which is 7 times more than the SRM which consists of 34 671 features\footnote{Note that we also tested a CNN with a feature vector of dimension similar to SRM, by keeping the pooling steps, and 64 filters per layer, and we obtained a probability of error 10\% better than with an EC with SRM.}.

After the two convolution layers, there is a fully connected three layer neural network. The first and the second layers consist of 1000 neurons, and the last layer only has 2 neurons. The operations carried out in the first and second layers are dot products, bias additions, and the applications of the ReLU activation function. The operations performed in the last layer are dot products, bias additions, and then a softmax in order to rescale the output values in $[0, 1]$.

\subsection{Clairvoyant scenario}
\label{ssec:clairvoyant}

Our first test is in the clairvoyant scenario. In this scenario, we put forward the hypothesis that the steganalyst knows the embedding algorithm, has good knowledge of the statistical distribution of the image database used by the steganograph, and knows the relative payload size. This scenario almost matches the Kerckhoffs principle. The steganalyst knows all the public parameters (in the clairvoyant scenario, the selection channel is assumed to be inaccessible), and does not know the private parameters such as the embedding secret key. This scenario is a laboratory scenario used to empirically assess the security of a steganographic embedding algorithm \cite{Ker2013_RealWorld}. We additionally make the hypothesis that the steganograph made the error to always use the same secret embedding key and thus that the steganalyst has access to "cover/stego" images pairs where stego images are generated with this same secret embedding key. Note that the embedding has been done with the simulator using always the same key.

Our tests were carried out on the cropped BOSSBase database, which consists of 40 000 grey-level images on 8 bits and whose size is $256\times256$. We embed with S-UNIWARD \cite{Holub2014} at 0.4 bits per pixels. The obtained set of images is thus made of 80 000 images (40 000 covers and 40 000 stegos). We limited our experiments to a single payload size due to the high number of computations, and the high number of experiments on the CNNs. 

\begin{figure}[!htbp]
  \centering
  \includegraphics[width=8cm]{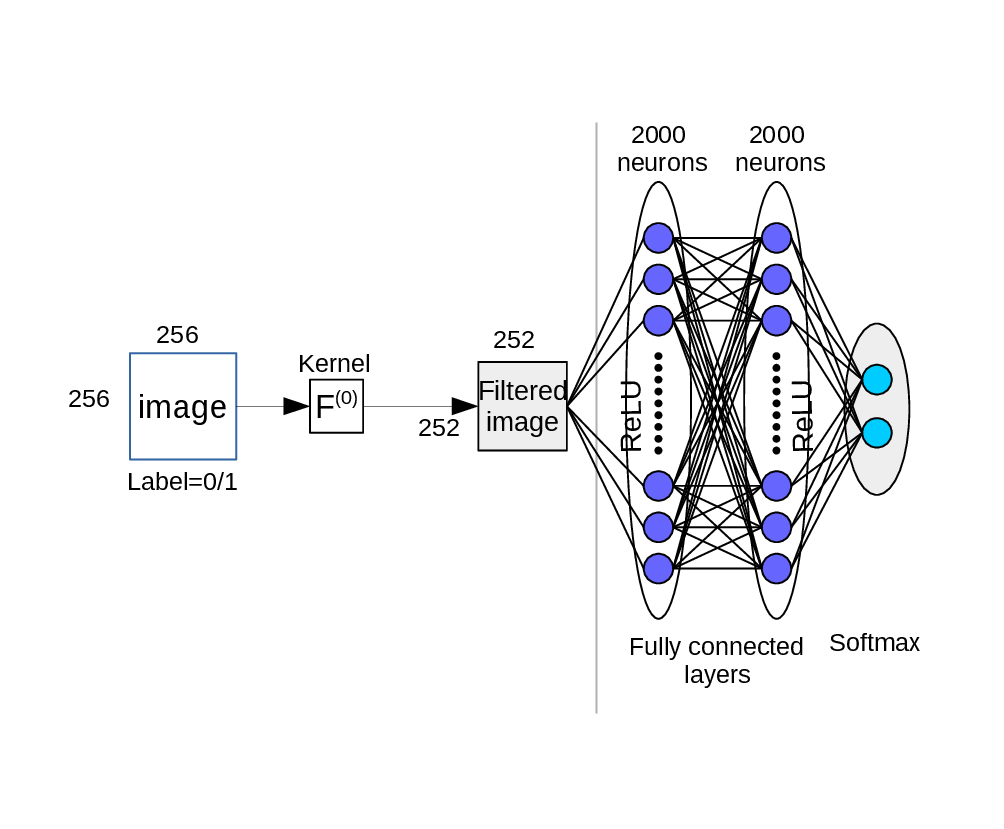}
  \caption{A Fully Connected Neural Network (FNN).}
  \label{fig:our_network_FNN}
\end{figure}

Three steganalysis approaches were evaluated. The first steganalysis was done using Rich Models \cite{Fridrich2012_Rich} and an Ensemble Classifier \cite{Kodovsky2012-EnsembleClassifiers}. The Rich Models is the SRM whose dimension is 34671. We will denote this steganalysis, {\bf RM+EC}, for Rich Models and Ensemble Classifier. The second steganalysis was done using the most efficient CNN we built (see Fig. \ref{fig:our_net}). We denote this steganalysis, {\bf CNN}, for Convolutional Neural Network. The third steganalysis was done using the a Fully Connected Neural Network (see Fig. \ref{fig:our_network_FNN}). We denote this steganalysis, {\bf FNN}, for Fully-Connected Neural Network. 

For each payload size, we conducted 10 tests where, for each test, the learning was done on 60 000 images randomly taken from the 80 000 images, such that the covers and stegos are always paired. The tests were performed on 10 000 images (5000 covers and 5000 associated stegos) randomly taken from the 20 000 remaining images. For the Ensemble Classifier, the decision threshold of each base learner was adjusted to minimize the total detection error under equal priors on the training set\footnote{We used two different Ensemble Classifiers. The one given on the Binghamton website\footnote{http://dde.binghamton.edu/download/}; in that case the learning set only consists of 10 000 images, and the one written in C++ and proposed in the paper \cite{Chaumont2012-EC-FS} with the features selection, and parallelization options; in that case, the learning set consists of 60 000 images. For those two cases, the results were similar, and we only report here the results obtained with the Binghamton code.}. For CNN, and FNN, the sum of the errors was minimized on the training set, as a gradient descent:
\begin{equation}
min \sum_{i=1}^{i=N_{TR}} \sum_{l=0}^{l=1} (h_l(x_i)-y_i)^2,
\end{equation}
with $N_{TR}$ being the number of training images, $h_0(.)$ (resp. $h_1(.)$), the CNN outputs giving a classification score for cover (resp. stego), $y_i \in\{0,1\}$ the label (0 for cover, and 1 for stego) for the $i^{th}$ input image $x_i$.

For both RM+EC, CNN, and FNN, the probability of error was computed and averaged over the 10 tests. The results are summarized in Table \ref{fig:result_claivoyant}. 

\begin{table}[!htbp]
\centering
\begin{tabular}{|c|c||c|c|}
 \hline
   & RM+EC & CNN & FNN\\
 \hline
   Max &  24.93\% & 7.94\% & 8.92\% \\
   Min & 24.21\% & 7.01\% & 8.44\% \\
   Variance & 0.14 & 0.12 & 0.16\\
 \hline
   Average & 24.67\% & 7.4\% & 8.66\%\\
 \hline
\end{tabular}
\caption{Table 1. Steganalysis results under the clairvoyant scenario for RM+EC, CNN, and FNN} \label{fig:result_claivoyant}
\end{table}

Rich Models with the Ensemble Classifier gives a probability of error of 24.67\%, whereas CNN gives a probability of error of only 7.4\%, and FNN gives a probability of error of 8.75\%. There is thus more than 17\% improvement by using our CNN, and more than 15\% with our FNN. This is an impressive improvement considering the difficulty to grab percentages on the probability of error in steganalysis. 

A first tentative explanation of the good behavior of the CNN has already been given in the previous Sections. Our CNN generates better results than Rich Models associated with an Ensemble Classifier for the following reasons:
\begin{enumerate}[label=\roman*)]
\item the shape of the CNN is well chosen, 
\item the learning process is done through a single global optimization, 
\item the filter kernels are optimized (which is not the case in Rich Models), 
\item there is a high number of filters (64 filters for the first layer) which enriches the diversity,
\item the second convolutional layer, which comes after what looks like a spatio-frequencial decomposition (obtained via the first layer), seems to act as the research of the presence of signals in all the spatio-frequential bands, 
\item we eliminated the pooling step (compared to the Qian {\it et al.} network) which was counter-productive because it was acting as a down-sampling, and thus leading to information loss.
\item the security error due to the lack of understanding from the steganograph about the security issues related to the use of key. Due to this security flaw, the embedding path is always the same and the embedding simulator always uses the same pseudo-random number sequence for generating the change probabilities. The steganalysis task is thus easier for a steganalyzer sensitive to the spatial content such as the CNN or the FNN (this also explain why a steganalyzer such as RM+EC, which is more sensitive to statistics, is less efficient).
\end{enumerate}
Note that a less structured network such as FNN gives impressive results even though it is less efficient than CNN. Note that the unknowns number for CNN is around 259 million (only 29 824 unknowns for the two convolutions layers, 
and 259 million for the fully connected layers 
), and for FNN is around 131 millions. 

We also carried out an additional experiment that confirmed that the first two CNN layers play a strong role in classification. The experiment first consisted of "cutting" another "smaller" CNN (with a pooling step) that was previously learned, by only keeping the two convolution layers. The network was then nothing more than a {\it feature extractor}; in this experiment there were only 57600 features. Second, this feature extractor was used to extract, on the same learning database, one feature vector per image. Third, an Ensemble Classifier used those feature vectors to learn a model. The probability of error was computed and averaged over 10 tests. We observed that the Ensemble Classifier, that learned with the features coming from the "smaller" CNN, allowed to reduce the average probability of error by 0.4\% w.r.t. the "smaller" CNN. This indicated that: i) feature extraction (i.e convolution steps) is the most interesting part of CNN, ii) the fully connected classifier part of CNN is not necessarily the best classifier. Note that this "smaller" CNN allowed us to obtain an improvement of 10\% compared to RM+EC.

We would also like to comment on the adaptive scenario, i.e adaptive steganalysis \cite{Tang2014_Adaptive}, also named selection-channel-aware steganalysis \cite{Denemark2014_AdaptiveSteganalysis}. When using adaptive steganalysis, the steganalyst uses an estimation of the modification probability of each pixel, as additional information for learning, in order to distinguish between cover and stego. In that scenario, authors from \cite{Denemark2014_AdaptiveSteganalysis} report a detectability improvement of 1 to 4\% with the Ensemble Classifier and SRM for detecting S-UNIWARD on the BossBase database. At $0.4$ bpp, the improvement was less than 2\%. Those results are not comparable with ours since our images are smaller, but the 2\% increment is really minor compared to the 17\% CNN increment. 

One should also mention the obtained results in the clairvoyant scenario when the steganograph uses a different key for each embedding. The probability of error of the CNN is 45.31\% (max=48.73\%, min=40.1\%, variance=0.001\%). The result is very bad and this is probably due to the fact that the CNN is not able to find any stego pattern. These result can be improved when increasing the numbers of filters. We experimented a test on a TitanX GPU card, with the Digits library, with a CNN of 128 filters for the first layer, 64 filters for the second one, 512 neurons on the first layer of the fully connected network, and 2048 for the second layer. The probability of error is reduced to 38.10\% (max=38.45\%, min=37.92\%, variance=0.000004). The increasing of the CNN size allows to improve the efficiency of the network but the performance is at the moment far from the state of the art (RM+EC).

\subsection{Cover-Source Mismatch scenario}

Our second test gave a very interesting results in the case of cover-source mismatch. The cover-source mismatch phenomenon occurs when the sources model, obtained during the learning step, differs from the sources that are used by the steganograph. From a geometrical point of view, we can explain this inconsistency problem by the fact that the cloud, describing the images used by the steganalyst, is not located at the same place as the cloud describing the images used by the steganograph. Only a few papers have assessed the cover-source mismatch in practice \cite{Pasquet2014}, and \cite{Lubenko_ker, Lubenko_mismatch}. Nevertheless, no satisfactory solution is currently available, even though there have been some attempts to understand the phenomenon \cite{Ker2014_Mismatch, Kodovsky2014_CoverSourceMismatch}. 

In this case, we subsampled the cropped BOSSBase five times, thus obtaining five different training sets. For each training set, we built a CNN, and then we applied the classification model on the LIRMMBase test database (The database is available at \url{http://www.lirmm.fr/~chaumont/LIRMMBase.html}). We report the average error probability over the five trials. Note that the cover-source mismatch was present since none of the BOSSBase cameras are in LIRMMBase. Note also that we used the same script as that used for BOSSBase in order to generate the grey-level images. Moreover, the original images we used to build the LIRMMBase, were uncompressed and came from three known databases: Dresden, Raise, and Columbia. The average probability of error, for RM+EC, CNN, and FNN are given in Table \ref{fig:result_csm}.

\begin{table}[!htbp]
\centering
\begin{tabular}{|c|c||c|c|}
 \hline
   & RM+EC & CNN & FNN\\
 \hline
   Max & 49.85\% & 5.90\% & 6.60\%\\
   Min & 47.20\% & 4.00 \% & 5.40\%\\
	 Variance &  0.35& 0.45 & 0.31 \\
 \hline
   Average & 48.29\% & 5.16\% & 5.89\%\\
 \hline
\end{tabular}
\caption{Table 2. Steganalysis results under the cover-source mismatch scenario for RM+EC, CNN, and FNN. The learning was done on BOSSBase and the tests on LIRMMBase.} \label{fig:result_csm}
\end{table}

We observed that the cover-source mismatch issue seriously affected the performance of the RM+EC classifier. Its results were close to those obtained by a random classifier. In return, CNN showed incredible robustness to the mismatch phenomenon with a $5.16\%$ probability of error. FNN gave similar results with a $5.96\%$ probability of error. Other experiments on BO\underline{\bf W}SBase \cite{BOWS2008} also confirmed that CNN is robust to the cover-source mismatch phenomenon, whereas RM+EC is not. BO\underline{\bf W}SBase is nevertheless not a very practical database since the set of used cameras is unknown, and since some of the cameras have also been used in BO\underline{\bf S}SBase. 


Surprisingly, the probability of error of CNN for LIRMMBase (5.16\%) was lower than that from the cropped BOSSBase (7.4\%). This was because there are fewer texture images in LIRMMBase than in the cropped BOSSBase. The LIRMMBase is thus easier to steganalyze. The CNN invariance to cover-source mismatch was so good that the mismatch phenomenon was no longer present, and we thus obtained steganalysis results that were more related to the content complexity of the data-base \cite{Cancelli2008} (Textured image databases are harder to steganalysis than homogenous ones).

The robustness was probably due to the transformation done by the two first layers (because keeping those layers and then branching an EC gives results similar to those of the whole CNN). Thus the feature extractor due to the first two layers gives features that are robust to the cover-source mismatch. Those features are probably invariant which means that the feature representation obtained after the second layer is not sensitive to the different image statistics. Assuming that the role of the first layer is to decompose the noise signal (if the image is stego, this noise is the stego noise) in a spatio-frequential decomposition, and that the role of the second layer is to detect the presence of particular patterns in the spatio-frequential bands, then this could explain the invariance to the image content and thus the invariance to the cover model. 

\begin{figure}[!htbp]
  \centering
  \includegraphics[width=5cm]{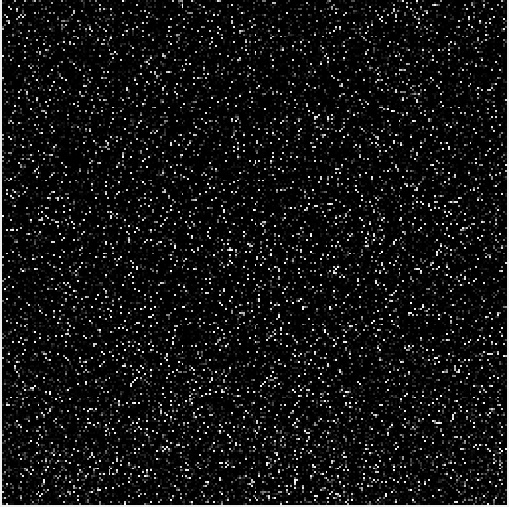}
  \caption{Embedding site. In white the most probable sites and in black the less probable ones.}
  \label{fig:embedding_site}
\end{figure}

Additionally, the robustness may also be explained by the fact that the CNN or the FNN are specialized in researching of a spatial pattern of {\it probability of change} strongly occurring in the "same embedding key" scenario. The Figure \ref{fig:embedding_site} illustrates the {\it probability of change} obtained after having counted how many times a pixel has changed for the stego images from the cropped BOSSBase database. One can observe that there are some pixels that never change in the stego images and some that change almost certainly. 

We also tested the cover-source mismatch scenario when the steganograph uses a different key for each embedding. The probability of error of the CNN is 42.07\% (max=43.50\%, min=40.90\%, variance=0.02\%). The result is slightly better than the RM+EC and is better than what was obtained in the cropped BOSSBase database (the result was 45.31\% for that scenario). 
With the CNN of 128 filters for the first layer, 64 filters for the second one, 512 neurons on the first layer of the fully connected network, and 2048 for the second layer, the probability of error is 48.99\% (max=49.06\%, min=48.92\%, variance=0.0000002) what is as bad as RM+EC.

\section{Further discussion}
\label{sec:discussion}

\subsection{Demystifying the CNN}

In order to demystify CNN, we already explained that the learning was equivalent to minimization of a function having many unknown parameters with a technique similar to gradient descent. In this subsection, we make links with previous research on the topic.

An important step of the CNN process is convolution as detailed in Sections \ref{ssec:convolution} and \ref{ssec:clairvoyant}. Learning of the filter kernels is done through minimization of the classification error using the back-propagation procedure. This is thus a simple optimization of filter kernels. This strategy already shown its efficiency in \cite{Holub2012_Filters}. In that paper, some of the filter kernel values, which were used for computing the feature vector, were obtained through optimization with the downhill simplex algorithm. The objective was to minimize the probability of error given by an Ensemble Classifier. The learning achieved in the convolution layers of a CNN shares the same idea that leads to customized kernels, well suited for steganalysis purpose.

When looking more precisely at the first layer, the kernel seems to act as a multi-band filtering (see Figure \ref{fig:filters_conv1}, where we can see the $64$ kernels of size $7\times7$ of our CNN). Some recent articles use such spatio-frequential decomposition (or a projection as in PSRM \cite{Holub2013_PSRM}) in order to compute Rich Models using Gabor filters \cite{Song2015_RichJPEGGabor} or DCT filters (DCTR features) \cite{Holub2015_DCTR}. Those filters are used to define projections that will then be used for computing a histogram leading to a feature vector.

When looking more precisely at the second convolution layer, which applies a very unusual convolution approach, nothing similar could be found in recent papers dealing with feature extraction, such as histogram computation \cite{Fridrich2012_Rich, Holub2013_PSRM}, non-uniform quantization \cite{Pevny2012_NonUniformQuantization}, feature selection \cite{Chaumont2012-EC-FS}, dimension reduction \cite{Pevny2013_ReductionDimensionCLS}, etc. Nevertheless something important is occurring in this second layer which allows us to obtain features unsensitive to the cover-source mismatch. By looking to the Equation \ref{eq:I2}, the role of this second layer looks like searching patterns in multiple bands. The sum of the convolutions from Equation \ref{eq:I2} is a way to accumulate presence clues of a searched signal in all the subands. The second layer then outputs a set of maps giving some clues to a signal presence.

To close the discussion about the CNN, we should also add that the last treatment of each layer is a normalization step, and that this type of processing can also be found in papers such as \cite{Kouider2013} or \cite{Cogranne2014_EC_HypothesisTest}. This normalization is done in order to obtain, comparable output values for each neuron. We should also mention that further analysis on the activation function is requiered to completely understand its impact. The activation introduced non-linearity, which is also the case in the Ensemble Classifier through the majority vote \cite{Kodovsky2012-EnsembleClassifiers}, or in Rich Models with the Min-Max features \cite{Fridrich2012_Rich}.

\subsection{"Same embedding key" scenario}

In order to conclude the discussion, we should mention papers of Ker \cite{Ker2008_location} and Quach \cite{Quach2014} that have studied the payload location estimation problem, with the hypothesis that the same embedding key is reused. Their scenario is different since it is used for the {\it forensic-steganalysis}. With the classical steganalysis, the steganalyst try to find if there is or not an embedded message.  With the {\it forensic-steganalysis}, the steganalyst often guess that embedding has occurred, and he tries to recover information such as the payload size, the embedding location, the secret keys, or the embedded message. 

Ker and Quach set their studies in the clairvoyant scenario (the steganalyst knows the embedding algorithm, has good knowledge of the statistical distribution of the image database used by the steganograph, and knows the relative payload size) and adds three additional hypothesis: (i) the steganograph always uses images of the same size, (ii) the steganograph always uses the same key, and (iii) the steganalyst has access to a set of stego images from the steganalyst, and he knows that those images are stego.

In the Ker's article \cite{Ker2008_location}, payload location is estimated for the LSB replacement algorithm with the use of the Weighted Stego-image (WS) steganalysis method \cite{Ker2008_WS} in order to estimate covers. With a set of stego images one can then accumulate clues about payload location. In the Quach proposition \cite{Quach2014}, payload location is estimated for the LSB replacement algorithm, the LSB matching algorithm, and the LSB Replacement Group-Parity Steganography. By using a Random Markov Model, Quach estimates the modified locations for each stego image from the test set, and then, by accumulating the location clues on all the test set, he can estimate sufficiently well the payload location (greater than 90\%) when the payload is high (0.5bpp) and the number of images is enough (100 to 10 000 depending of the steganalyzed algorithm). 

Independently from the scenario (classical steganoganalysis or {\it forensic-steganalysis} scenario), and independently from the use of adaptive or non-adaptive steganography, we can observe that those articles and our article, use the security weak due to the error made by the steganograph, and consisting to reuse the same key. As stated before, this error leads to an easier localization of suspect/modified pixels because the {\it spatial pattern of probability of change} can be learned such as with our approach, or can be recovered as in papers \cite{Ker2008_location} and \cite{Quach2014} by accumulating clues obtained from a set of stego images. With the knowledge of this {\it spatial pattern of probability of change} it becomes easier to apply a classical steganalysis, as in the present article, or a {\it forensic-steganalysis} such as payload localization as in \cite{Ker2008_location} and \cite{Quach2014}.

\section{Conclusion}

In this article, we propose to pursue the study of CNNs for steganalysis. We tested more than 40 CNNs and found the right parameters for the steganalysis domain in the scenario where the steganograph reuses the same secret embedding key for embedding with the simulator in images. Instead of using a very deep network, such as that proposed by Qian {\it et al.}, our experiments led us to use a network that is in height, and only consists two convolutional layers. We replaced the unconventional Gaussian activation function by a more classical ReLU activation function, we suppressed the pooling step that was acting as a down-sampling, and we pre-processed the images by applying high pass filtering before feeding the CNN.

We evaluated CNNs in two different scenarios. The first set of tests was done with the clairvoyant scenario. We used the cropped BOSSBase database with embedding with S-UNIWARD at $0.4$ bpp. Compared to the state-of-the-art approach, i.e. the Ensemble Classifier with SRM features, CNN and FNN reduce the classification error by a three fold.

The second set of tests was done with the cover-source mismatch scenario. We used the BOSSBase with S-UNIWARD embedding at $0.4$ bpp for learning, and tests were carried out on the public LIRMMBase database. The cover-source mismatch was fully achieved since the cameras were different from one base to another. The conventional method (RM+EC) totally failed to detect the use of steganography in LIRMMBase since the classification error was 48.29\% i.e almost a random classification. Conversely, CNN exhibited natural invariance to the cover-source mismatch with a classification error of 5.16\%.

Our future studies will concentrate on different scenarios, on improving some of the network parameters, and on gaining insight into the network behavior. Moreover, further experiments have to be done with different payload sizes, and different algorithms.



\section{Acknowledgments} 
The authors thank Jessica Fridrich and her team for the numerous discussions about this article.



{
\bibliographystyle{IEEEbib}
\bibliography{strings,bib}
}

\begin{biography}
\begin{minipage}[b]{0.25\linewidth}
\noindent \hspace{0.4cm}\includegraphics[width=2cm]{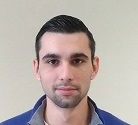}
\end{minipage}
\vspace{0.1cm}
\begin{minipage}[b]{0.65\linewidth}
Lionel PIBRE received his Master's degree in Computer Science in 2015 from the University of Montpellier, France. He is currently working toward the Ph.D. degree in the LIRMM laboratory of Montpellier. His 
\end{minipage}
\begin{minipage}{\linewidth}
research interests are steganography / steganalysis, urban objects detection, and segmentation in aerial photography.
\end{minipage}

\vspace{0.5cm}

\begin{minipage}{0.25\linewidth}
\hspace{-0.2cm}\includegraphics[width=2cm]{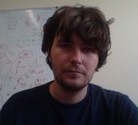}
\end{minipage}
\vspace{0.1cm}
\begin{minipage}{0.65\linewidth}
J\'er\^ome PASQUET received his Master's degree in Computer Science in 2013 from the University of Montpellier, France. He is currently working toward the Ph.D. degree in the LIRMM laboratory of Montpel-
\end{minipage}
\begin{minipage}{\linewidth}
lier. His research interests are steganography / steganalysis, urban objects detection, and segmentation in aerial photography.
\end{minipage}

\vspace{0.5cm}

\begin{minipage}{0.25\linewidth}
\hspace{-0.2cm}\includegraphics[width=2cm]{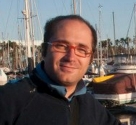}
\end{minipage}
\vspace{0.1cm}
\begin{minipage}{0.65\linewidth}
Dino IENCO received his PhD in Computer Science in 2010 from University of Torino. From 2011 he is researcher at at the Irstea Institute, Montpellier, France. His main topics of research involved clus-
\end{minipage}
\vspace{0.1cm}
\begin{minipage}{\linewidth}
tering algorithm for textual information, social network analysis, data mining approaches for biological data, supervised and semi-supervised methods for multimedia data (image and document analysis) and graph mining. He is actively working in the field of spatio-temporal data with a major emphasis on remote sensing analysis. In 2009, he visited the Yahoo! Research Lab in Barcelona (Spain) to work on information propagation. He co-authored more than 40 papers in major international conferences (SDM, ECML/PKDD, EMNLP, ECIR, etc.), and journals (DAMI, ACM TKDD, Pattern Recognition, JIIS, etc.).
\end{minipage}

\vspace{0.5cm}

\begin{minipage}{0.25\linewidth}
\hspace{-0.2cm}\includegraphics[width=2cm]{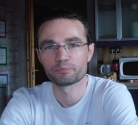}
\end{minipage}
\vspace{0.1cm}
\begin{minipage}{0.65\linewidth}
Marc CHAUMONT received his Engineer Diploma in Computer Sciences at the INSA 
of Rennes, France in 1999, his Ph.D. 
at the IRISA Rennes 
in 2003, and his HDR ("Habilitation \`a Diriger des Recherches") 
\end{minipage}
\begin{minipage}{\linewidth}
at the University of Montpellier 
in 2013. Since September 2005, he is an Assistant Professor in the LIRMM laboratory
of Montpellier and the University of Nîmes. His research areas are multimedia security (steganography, watermarking, digital forensics, video \& image compression) and segmentation \& tracking in images and videos. He is member of the TC of IEEE SPS - Information Forensics and Security for the period 2015-2017. He was program chair of ACM IH\&MMSec'2013. He is reviewer for more than 20 journals (IEEE TIFS, IS\&T JEI, ...) and for more than 10 conferences (EI MWSF, IEEE WIFS, ACM IH\&MMSec, IEEE ICIP, ...).
\end{minipage}

\end{biography}

\end{document}